\documentclass[final,5p,times,twocolumn]{elsarticle}

\usepackage{hyperref}

\usepackage{amsmath,amssymb}
\usepackage{graphicx,tabularx}
\usepackage{epsf}       
\usepackage{epsfig}       
\usepackage[ngerman, english]{babel}
\usepackage{placeins}
\usepackage{siunitx}
\usepackage{upgreek}
\usepackage{setspace}
\usepackage{bold-extra}
\usepackage{comment}
\usepackage{multicol}
\usepackage{float}
\usepackage{afterpage}
\usepackage{eurosym}
\usepackage[hang]{footmisc}
\usepackage{lipsum}
\usepackage{subcaption}
\usepackage{color}

\journal{Nuclear Instruments and Methods in Physics Research A}

\begin{document}
\begin{frontmatter}

\title{Characterization of Passive CMOS Strip Sensors\tnoteref{mytitlenote}}


\author[mymainaddress]{Leena Diehl\corref{mycorrespondingauthor}}
\cortext[mycorrespondingauthor]{Corresponding author}
\author[myverylastaddress]{Marta Baselga}
\author[mysecondaryaddress,bonn]{Ingrid Maria Gregor}
\author[mymainaddress]{Marc Hauser}
\author[bonn]{Tomasz Hemperek}
\author[mymainaddress]{Jan Cedric Hönig}
\author[mymainaddress]{Karl Jakobs}
\author[mymainaddress]{Sven Mägdefessel}
\author[mymainaddress]{Ulrich Parzefall }

\author[mymainaddress]{Arturo Rodriguez}
\author[mysecondaryaddress]{Surabhi Sharma}
\author[mymainaddress]{Dennis Sperlich}
\author[mymainaddress]{Liv Wiik-Fuchs}
\author[lastaddress]{Tianyang Wang}

\address[mymainaddress]{Physikalisches Institut, Albert-Ludwigs-Universität Freiburg, Hermann-Herder-Straße 3, Freiburg, Germany}
\address[mysecondaryaddress]{Deutsches Elektronen Synchrotron DESY, Notkestr. 85, Hamburg, Germany}
\address[bonn]{Physikalisches Institut, University of Bonn, Nussallee 12, 53115 Bonn, Germany}
\address[lastaddress]{Zhangjiang Laboratory, No. 99 Haike Road, Zhangjiang Hi-tech Park, Pudong, Shanghai, P.R.China}
\address[myverylastaddress]{Physik E4, TU Dortmund, Otto-Hahn-Strasse 4a, 44227 Dortmund, Germany} 
\begin{abstract}
Recent advances in CMOS imaging sensor technology , e.g. in CMOS pixel sensors, have proven that the CMOS process is radiation tolerant enough to cope with certain radiation levels required for tracking layers in hadron collider experiments. With the ever-increasing area covered by silicon tracking detectors cost effective alternatives to the current silicon sensors and more integrated designs are desirable.  
This article describes results obtained from laboratory measurements of silicon strip sensors produced in a passive p-CMOS process. Electrical characterization and charge collection measurements with a $^{90}$Sr source and a laser with infrared wavelength showed no effect of the stitching process on the performance of the sensor.  

\end{abstract}


\end{frontmatter}


\section{Introduction}
All current and envisaged high energy physics detectors rely on silicon-based tracking systems due to their superior resolution, high radiation tolerance and readout speed.
Sensors fabricated in CMOS processes offer an interesting alternative to the current sensor technology, with the possibility to combine the active detection layer and readout electronics into a single structure. 
This integration allows for thinner sensors, reducing the amount of material the particles have to pass through, and for pixel sensors a small pixel size is possible, enlarging the granularity. However, the passive CMOS sensors investigated misuse the CMOS process used to produce CMOS integrated circuits, without including any readout circuit. 
Furthermore, the CMOS process is the industrial standard process, it allows for a large production volume at a rather low cost. 

However,  the suitability of the passive CMOS process for strip sensors needs to be evaluated. 
Upcoming experiments will need to cover large areas with silicon sensors, e.g. more than $170~\mathrm{m}^2$ for the ATLAS Inner Tracker \citep{CERN-LHCC-2017-005}, thus sensors of a size exceeding the commercially used reticles will be necessary. These masks contain the pattern of a single layer of the sensors and usually have a size of $1-4~\mathrm{cm}^2$ . 
Therefore, larger structures have to be realized with a process called stitching \cite{turchetta2011large}. The method divides the sensor into smaller areas, fabricated separately with a specific mask, which are then connected. In this case, it is done using a reticle divided into six parts, as described in \cite{bryant20021024x1024}. Stitching increases the price about 30\% compared to a non-stitched device, however for large volumes the ability to use commercial production lines most likely still makes the sensors cheaper than other comparable sensors.

The sensors in this study were produced by LFoundry \cite{lfoundry} and are cut from a wafer with a resistivity of $3-5~\mathrm{k}\Omega\mathrm{cm}$  produced with the Float-Zone technique. 
The sensors are fabricated with a $150~\mathrm{nm}$ process on a p-type bulk and have a nominal thickness of $(150\pm15)~\upmu\mathrm{m}$. To the thinned backside of the sensors a $p^+$ implant is implemented, before it is annealed with a laser and a metal layer is added at an external company. The sensors from the first batch do not have a metal layer, and the doping concentration of the $p^+$ layer was comparatively low.  The studies are performed on sensors in two sizes, both types are one cm broad and they have a strip length of  $2.1~\mathrm{cm}$ or $4.1~\mathrm{cm}$, referred to as short and long sensors, respectively.

\begin{figure}[ht]
    \centering
    \includegraphics[width=0.95\linewidth]{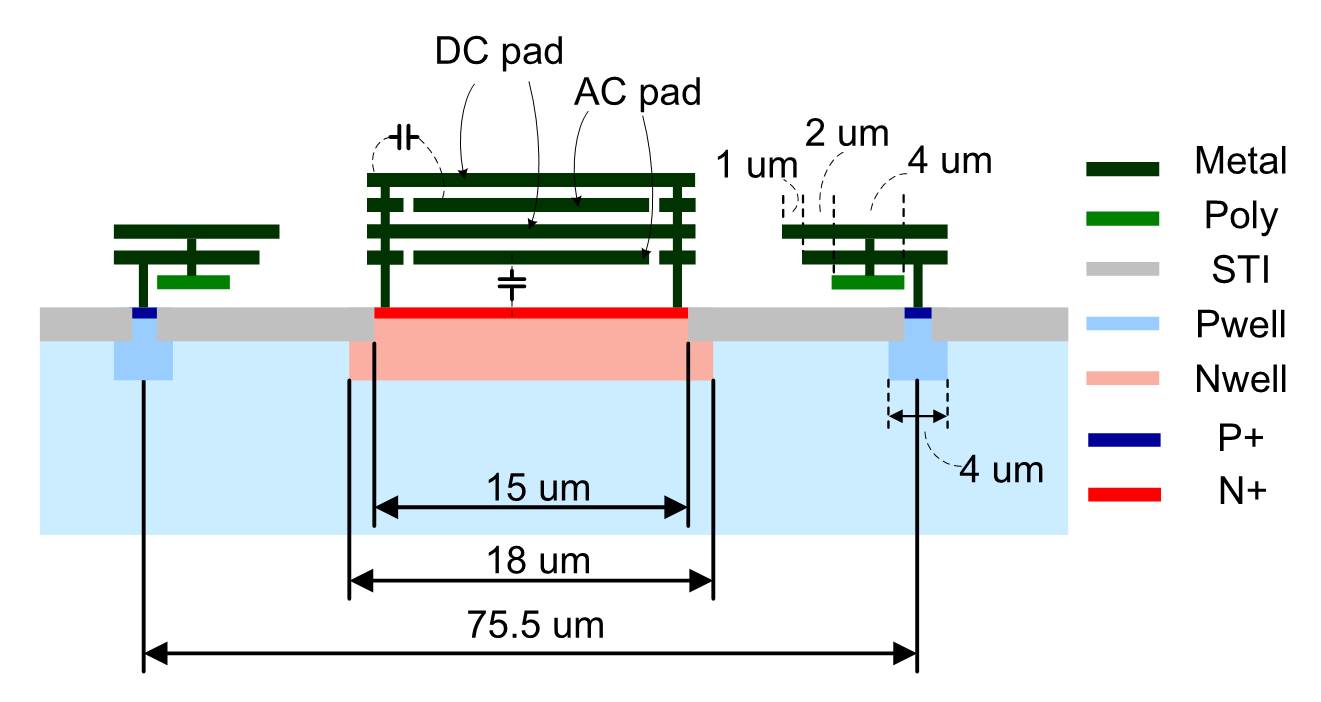}
    \caption{\it \small Regular implant strip design. 
    \label{fig:regular}}
\end{figure}

\begin{figure}[ht]
    \centering
    \includegraphics[width=0.95\linewidth]{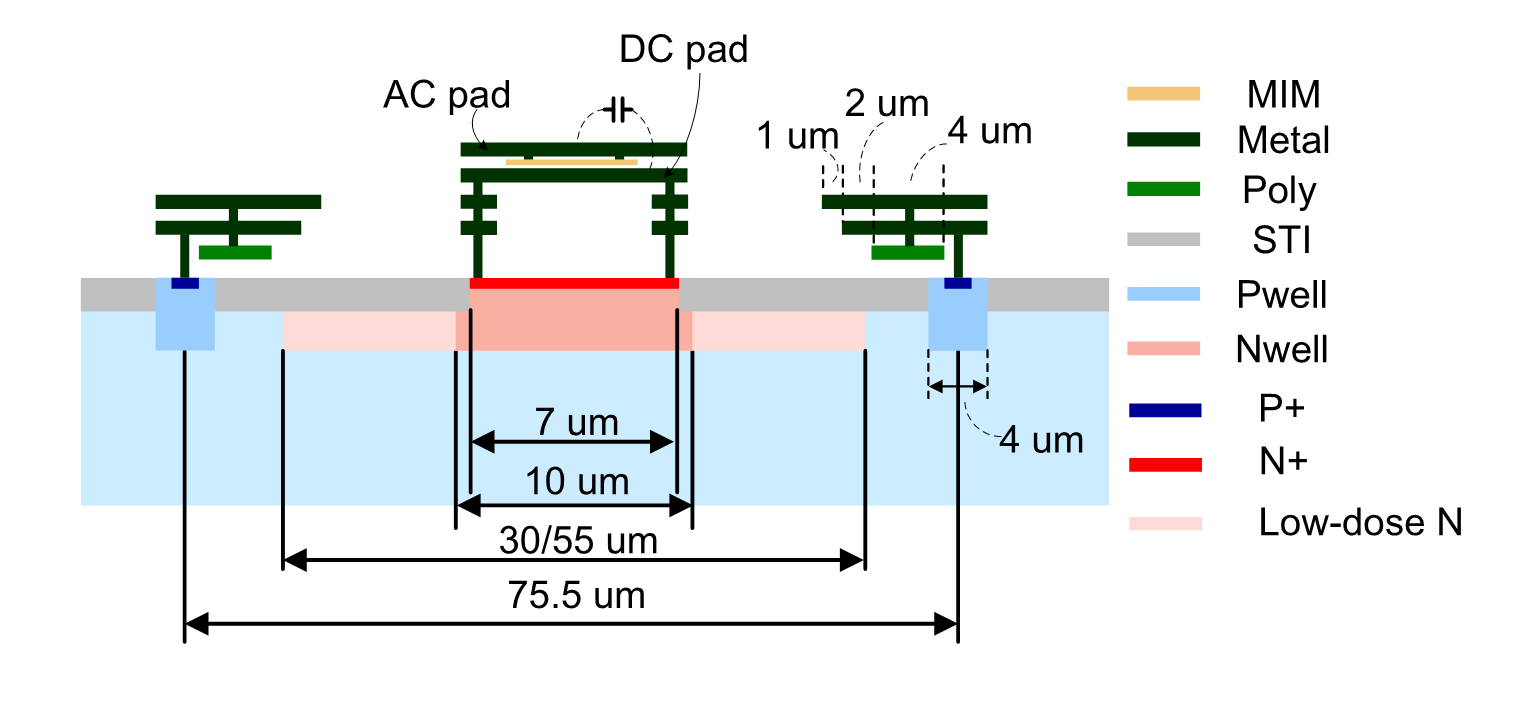}
    \caption{\it \small Low dose implant strip design with either 30 or 55~$\upmu\mathrm{m}$ n-well width.
    \label{fig:low dose}}
\end{figure}

The sensors are split into three designs to investigate different depletion concepts by varying the doping concentration of the $n$-well as well as its width. The $n$-well was chosen as it was known to work well from previous works with monolithic chips. Compared are the \textit{regular} design, as described in Figure \ref{fig:regular}, and the\textit{ low dose} design (see Figure \ref{fig:low dose}), which comes additionally in two different $n$-well implant widths, $30~\upmu\mathrm{m}$ and $55~\upmu\mathrm{m}$.  The designs differ in the implant width of the high dose $n^+$ region as well as the low dose $n$ region. They are not the standard $n$-well design of LFoundry, the implant designs were evaluated with TCAD simulations and the parameters were given to the company in order to test their effect on the electric field distributions and on the radiation hardness of the sensors. All designs have a metallization with an AC and a DC pad on the strip implant, which slightly differ in the layout. The \textit{ regular} design with the broader strip implants uses four metal layers, where the AC pads are coupled with a capacitor, while the \textit{low dose} design uses a metal-insulator-metal (MIM) capacitor to separate the two metal layers of the AC and the DC pad. Thus, the AC coupling is done using CMOS capacitors instead of the $SiO_2$-layer commonly used for silicon sensors. The implemented $p$-well structure, which includes a metallization as well, is the same for all designs.

\begin{figure}[ht]
    \centering
    \includegraphics[width=\linewidth]{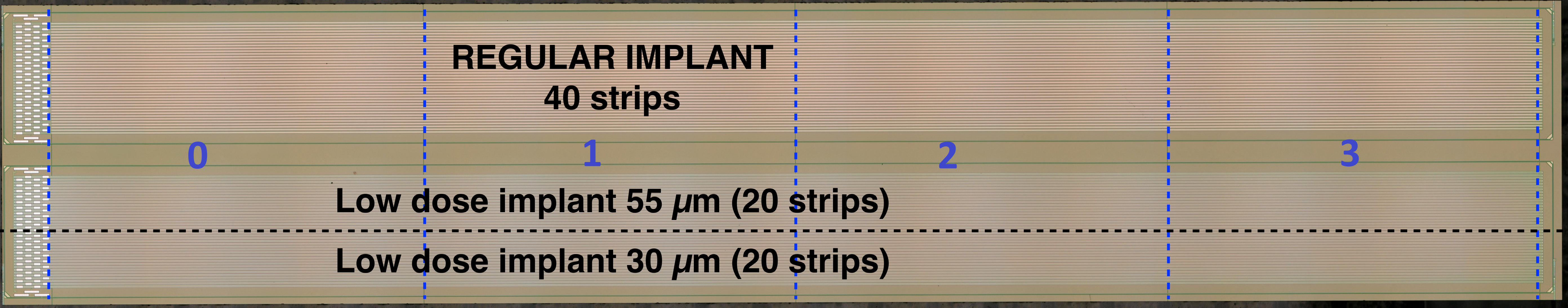}
    \caption{\it \small \SI{4}{cm} long strip sensor. The blue dashed lines show the stitching positions, the investigated stitches are labelled 0-3. 
    \label{fig:long_picture}}
\end{figure}

How the three different designs are distributed in the sensors is shown in Figure \ref{fig:long_picture} on a microscope picture of a long sensor, where the dashed blue lines additionally indicate the stitch lines. The further used labelling of the investigated stitches is sketched as well. 
The \textit{regular} design includes 40 strips on the upper half of the sensor and has its own bias ring, while the \textit{low dose 55} (middle) and \textit{low dose 30} (bottom) designs have 20 strips each and share one bias ring. All strips have a pitch of $75.5~\upmu\mathrm{m}$ and are connected via a bias resistor to the bias rings. Due to the two separate bias rings, the\textit{ regular} half and the \textit{low dose} half can be biased either individually, or together by connecting them via wirebonds. However, the two different designs in the \textit{low dose} half cannot be bias separately, thus the electrical characterizations differentiate only between \textit{regular} and \textit{low dose} design. Outside of the bias rings, there are five guard rings to shape the electric field at the edge of the sensor and prevent an early breakdown.

\section{Sensor Characterisation}

To test the sensors before any further studies, all sensors that are part of this study were electrically characterised.  The current-voltage (IV) and capacitance-voltage (CV) behaviour were measured on a probe-station at room temperature, using needles to apply the bias voltage on the sensor.

\begin{figure}[h]
    \centering
    \includegraphics[width=0.7\linewidth]{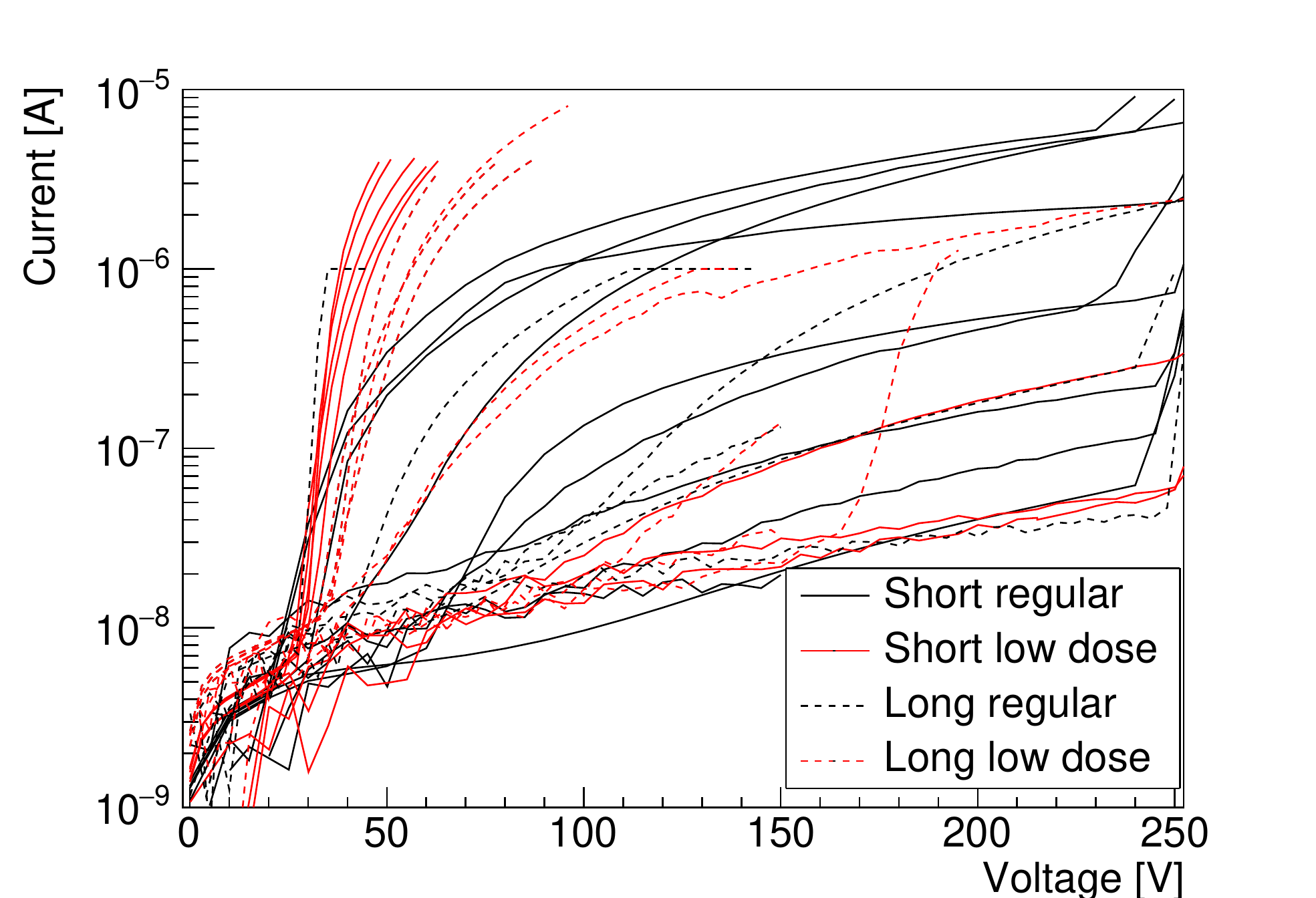}
    \caption{\it \small IV characteristics of the strip detectors. 
    \label{fig:iv}}
\end{figure}

Full depletion and breakdown voltage were investigated with IV-measurements for all sensors. The needle is placed on the bias ring, connected to all strips, while the high voltage is applied on the backplane, the measurements are shown in Figure \ref{fig:iv}, for all available sensors. 

The sensors show a rather high leakage current and a breakdown before \SI{250}{\V}. Some sensors even break down at around \SI{40}{\V}, when the depletion zone reaches the backplane. This is most likely caused by a non-optimal $p^+$ implant at the backplane due to a too low implant dose for the first production batch, and a missing homogeneous metal layer. The sensors show a broad spread in measured current and breakdown voltage. They were found to react rather sensitive to handling, e.g. mechanical pressure, small scratches and humidity, therefore the IV-characteristics are not uniform, but depend strongly on the individual sensor.  

\begin{figure}[h]
    \centering
    \includegraphics[width=0.7\linewidth]{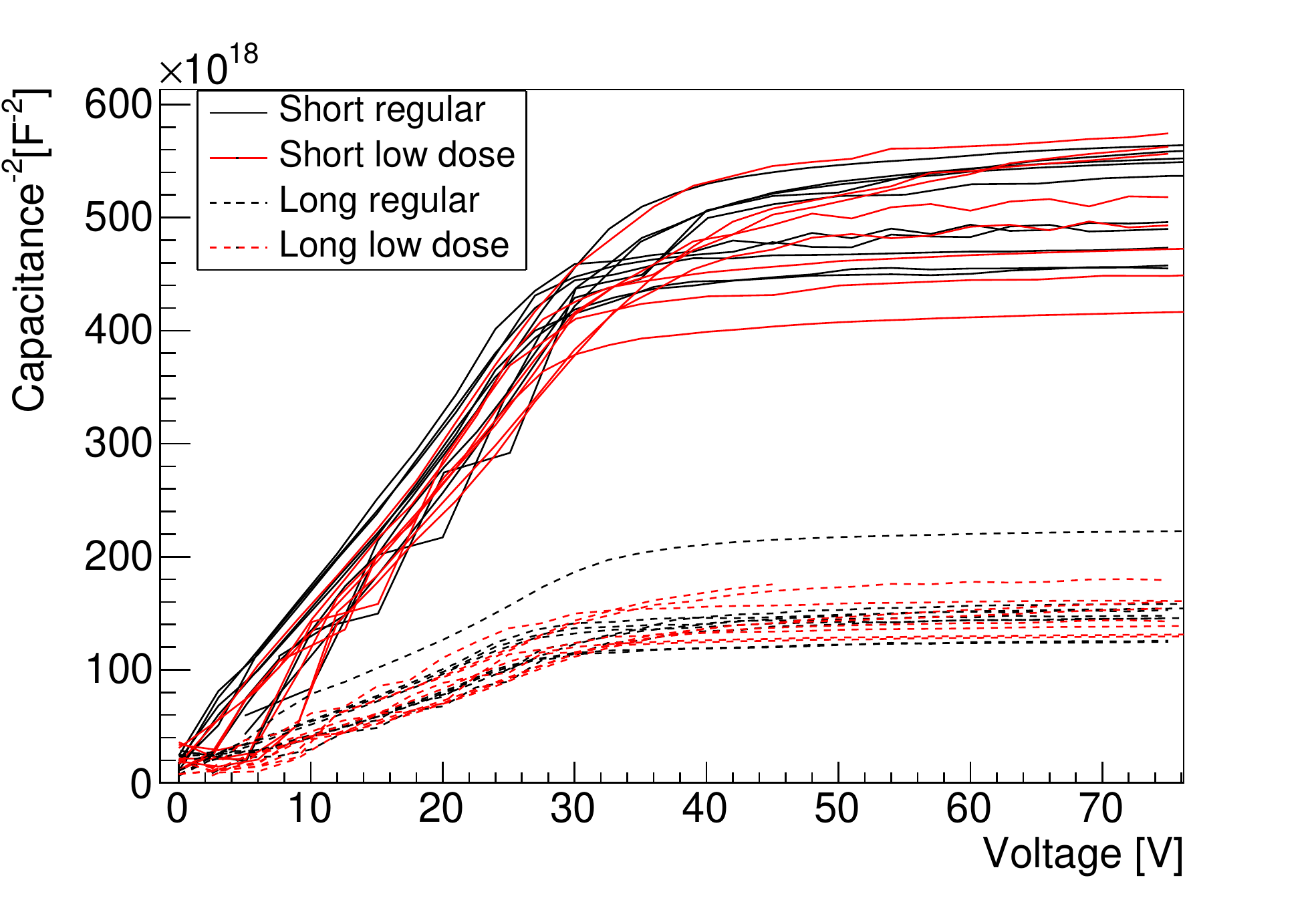}
    \caption{\it \small CV characteristics of the strip detectors. 
    \label{fig:cv}}
\end{figure}

 Figure \ref{fig:cv} shows the CV characteristics measured at a frequency of \SI{1}{\kilo\hertz}. The sensors show a depletion voltage around \SI{30}-\SI{35}~V, regular strips show a slightly lower depletion voltage than the low dose ones.  
As expected due to the smaller sensor area, the bulk capacitance of $50~\mathrm{pF}$ of the short sensors is lower than the $100~\mathrm{pF}$  measured for the long sensors. 

\section{TCAD Simulations}

\begin{figure}[h]
    \centering
    \includegraphics[width=0.88\linewidth]{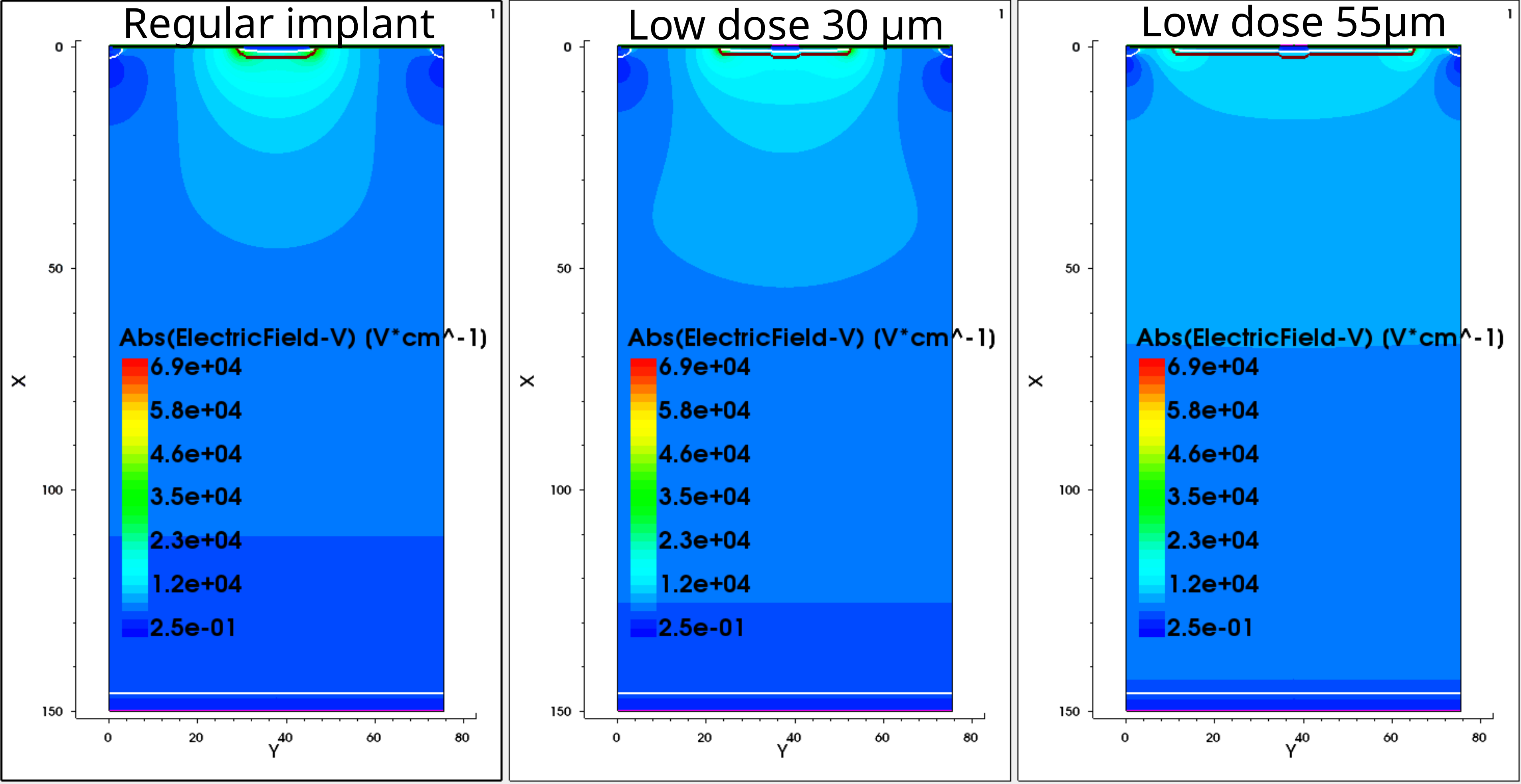}
    \caption{\it \small Simulation of the electric field at \SI{100}{\V} for the three different geometries. 
    \label{fig:simulation_E}}
\end{figure}

To better understand the different geometries, TCAD simulations (Technology Computer-Aided Design) with Synopsys Sentaurus \cite{synopsys} were carried out using the three strip designs: \textit{regular} implant, \textit{low dose 30}~${\upmu\mathrm{m}}$ implant and \textit{low dose 55}~${\upmu\mathrm{m}}$ implant. The simulations took into account a $150~\upmu\mathrm{m}$ thick sensor with  a $75.5~\upmu\mathrm{m}$ strip pitch, considering only two dimensions ($1~\upmu\mathrm{m}$ of strip length), with the implants distances shown in Figures \ref{fig:regular} and \ref{fig:low dose}. A standard $p^{++}$ implant and metallisation layer was assumed for the backplane in the simulation. For all simulations, the sensor is biased from the backplane and  \SI{0}{\V} are applied on the $n^{++}$ implant.
Figure \ref{fig:simulation_E} shows the electric field at \SI{100}{\V} for the three geometries. They show different electric field shapes, the narrow implants (\textit{regular}) show also narrower electric field lines, while the wider implants (\textit{low dose 55}$~\upmu\mathrm{m}$) show a higher electric field spread through the entire strip width.

\begin{figure}
    \centering
    \includegraphics[width=0.6\linewidth]{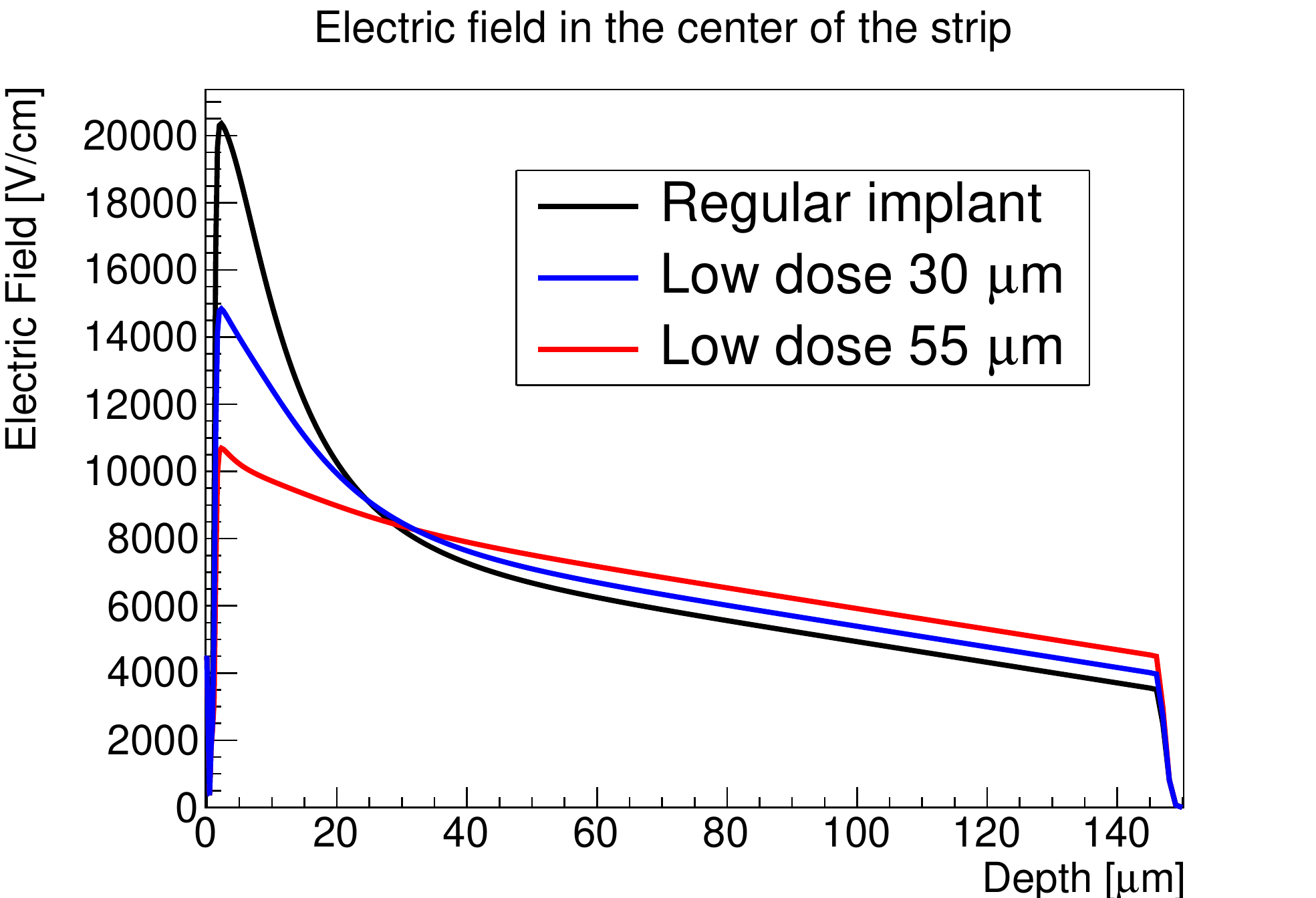}
    \caption{\it \small Simulation of the electric field cut at the center of the strip applying \SI{100}{\V} of bias voltage for the three different geometries. 
    \label{fig:simulation_E_cross}}
\end{figure}

Figure \ref{fig:simulation_E_cross} presents the electric field in a vertical cut through the center of the strip to the backplane. The \textit{regular} implant shows a higher electric field at the center of the strip than the \textit{low dose} implant due to the different strip designs, especially the different implant widths.  

\FloatBarrier

\begin{figure}
    \centering
    \includegraphics[width=0.6\linewidth]{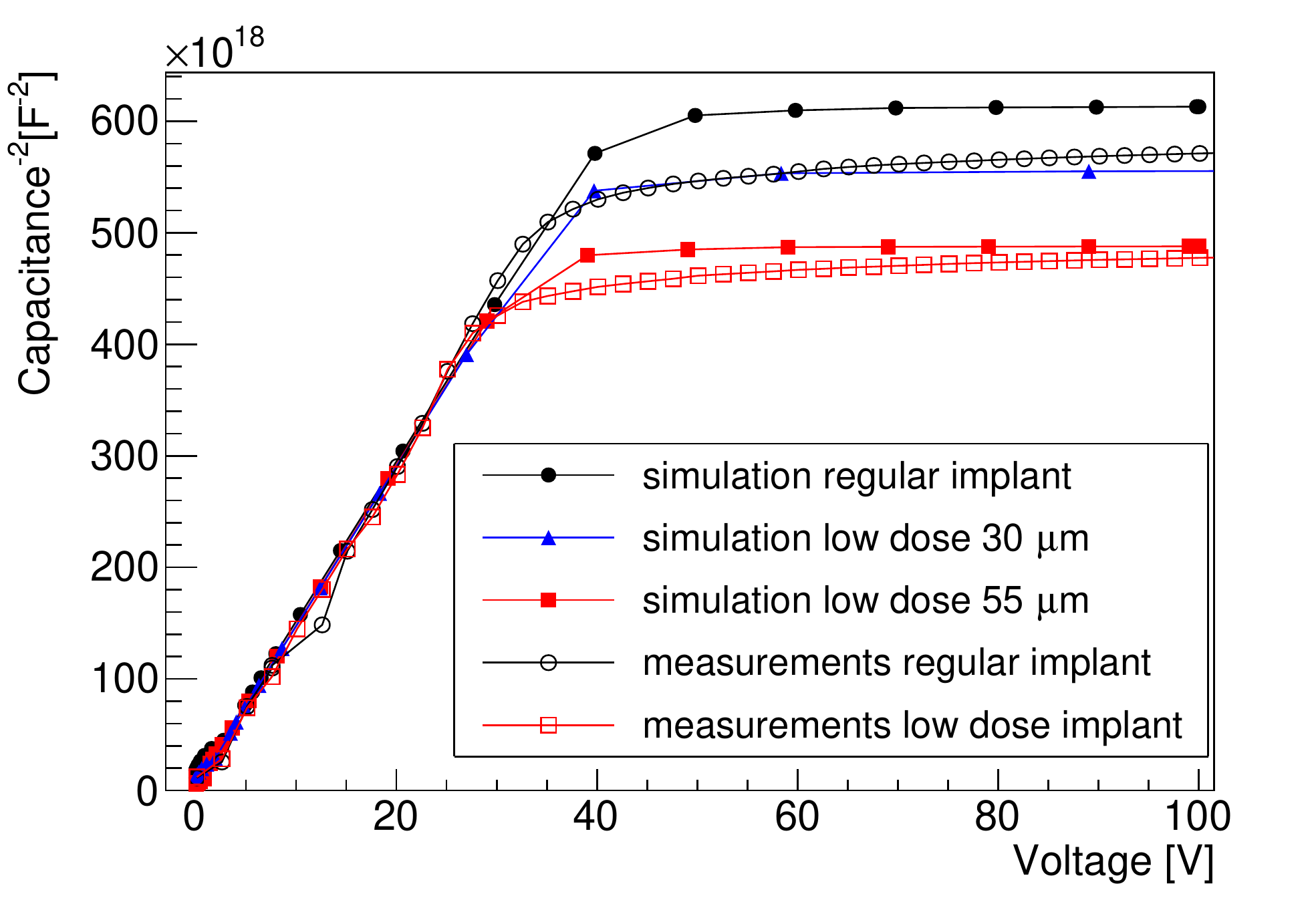}
    \caption{\it \small CV characteristics of the short strip detectors and simulation for the three geometries.
    \label{fig:cv_sim}}
\end{figure}

Figure \ref{fig:cv_sim} shows the simulation of the CV curves for the three different geometries of a short sensor, compared to one measurement of a short sensor from the \textit{regular} and one from the \textit{low dose} design. The measurements separate only between \textit{regular} and \textit{low dose} design since the two low dose implants cannot be measured separately from the bias ring. The regular implant yields a lower capacitance than the low dose implant geometries, showing the agreement with the simulations and measurements. The slight shift observed between simulations and the experimental data is due to uncertainties in the doping profile values used for the simulations.

\section{Source Measurements}
\label{sourcemeasuremnts}
Charge collection measurements were performed using Minimal Ionizing Particle (MIP)-like electrons originating from a $^{90}\mathrm{Sr}$ source, triggered by two scintillators behind the sensors and therefore rejecting low-energy particles.

The ALIBAVA system \cite{MarcoHernndez2010} is used for data acquisition. The sensors are connected to a 128-channel Beetle ASIC \cite{Lochner:2006vba}, the pulses of each channel are sampled at a frequency of 40~MHz and the output is stored in an analogue pipeline. The data is converted to digital counts with a 10-bit analogue to digital converter, the gain of the ASICs is temperature dependent and varies for each ASIC.
Measurements were taken at different bias voltages ranging from $5 - 60$~V and the signals were analysed using a customised software package, including iterative common mode and pedestal subtraction algorithms. Additionally, channels exhibiting an excessive noise behaviour were masked. 
The charge is determined using a clustering algorithm with a seed cut of $3.5$ and a neighbouring cut of $1.8$ times the noise value, to exclude spurious hits and noise contributions to the signal. 

\begin{figure}[ht]
    \centering
    \includegraphics[width=0.75\linewidth]{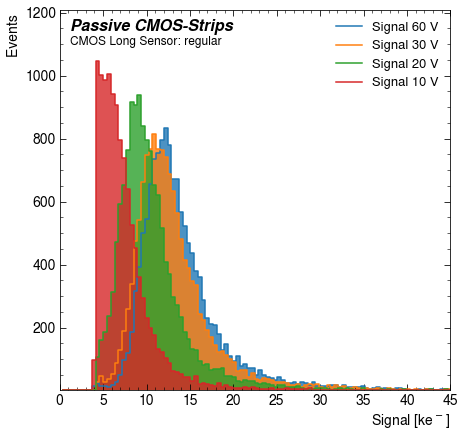}
    \caption{\it \small Number of events as a function of the measured signal of the regular design of a long sensor for several voltages.
    \label{fig:Landauas_longSensor}}
\end{figure}

In order to evaluate the sensor performance especially in regard to the stitching process, different measurements were conducted with the source pointing at the individual stitches only, where the segments are numbered starting with zero. This was performed for the three different designs, \textit{regular}, \textit{low dose 30} and \textit{low dose 55}.

The collected charge is defined as the most probable value (MPV) of a Landau-Gauss fit to the signal distribution. 
The signal distribution for four different voltages measured in the \textit{regular }design of a long sensor is displayed in Figure~\ref{fig:Landauas_longSensor}, where the number of events is plotted as a function of the measured signal in k$e^-$. 
The signal distribution shifts more towards higher values with increasing voltage, as expected due to the increase of the depleted area, however the integral of all events remains approximately the same for all voltages. The shape of the signal measured at 10~V varies the most from the others, since there some signals are very low and fall below the threshold, thus the distribution is cut off. 

\begin{figure}[ht]
    \centering
    \includegraphics[width=0.75\linewidth]{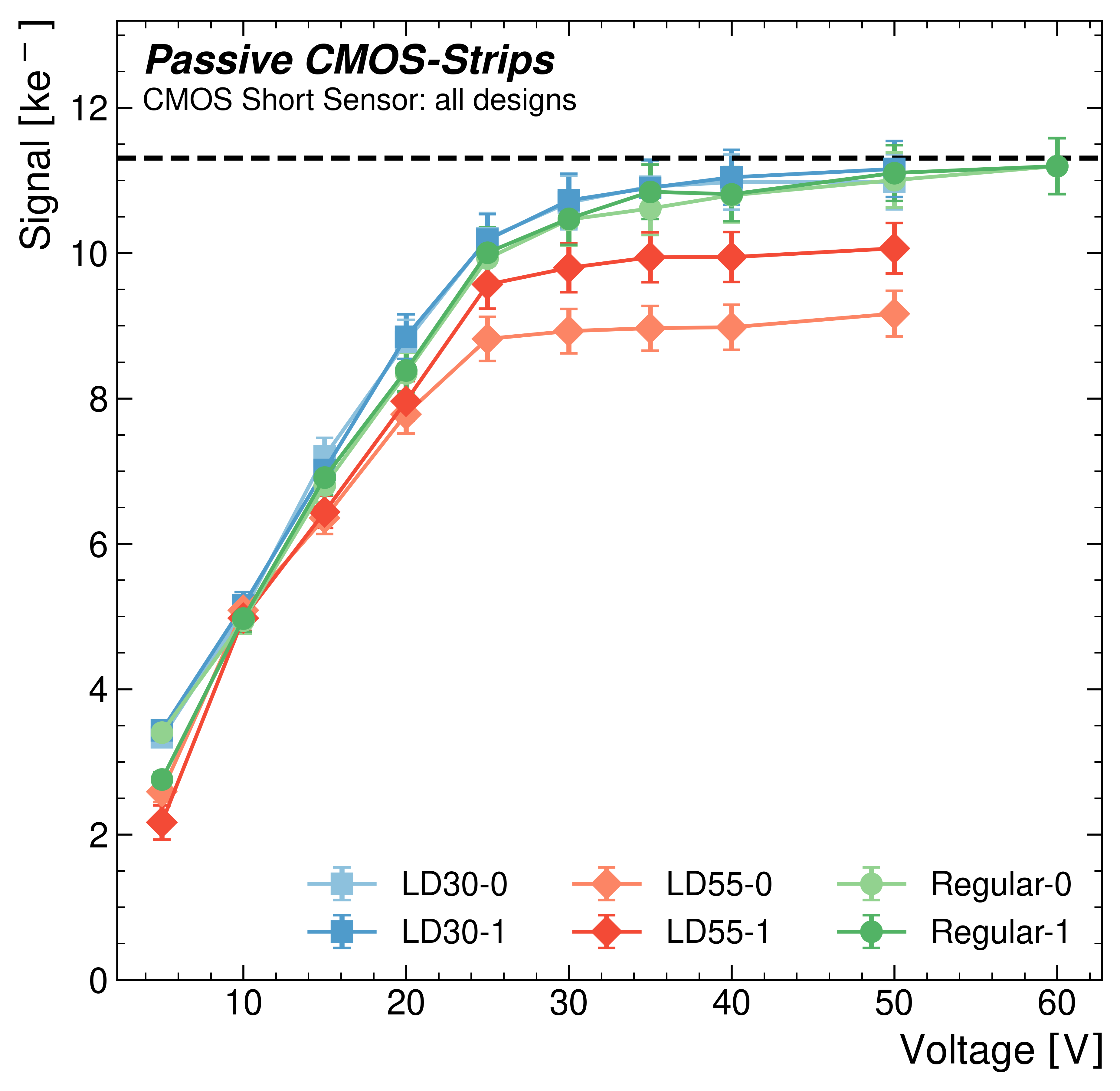}
    \caption{\it\small Signal as a function of the bias voltage in different stitched areas of a short CMOS strip sensor. The stitching areas are numbered 0 and 1 as indicated in the label names (e.g. LD30-0 means \textit{low dose 30} stitch 0). The black dashed line acts as eye-guide for the approximately expected charge.)
    \label{fig:short sensor signal}}
\end{figure}

The collected charge as a function of the voltage is shown in Figure~\ref{fig:short sensor signal} for a short sensor for the different designs and stitched areas. The dashed black line at $11.31~\mathrm{k}e^-$ indicates the expected amount of charge collected in a fully depleted \SI{150}{\micro\m} thick silicon sensor for a MIP-like particle, however, it is only an eye-guide. The fluctuations on the active thickness as well as systematic errors in the calibration of the readout board can affect the amount of collected charge. As can be seen, the sensor reaches full depletion between 30 and 35~V, and the \textit{regular} and the \textit{low dose 30} designs reach the expected amount of charge within the error bars.
The slightly lower collected charge of the \textit{low dose 55} design is explained with the small number of channels available for analysis. Since there are only 20 channels for each of the \textit{low dose} designs, and several noisy channels had to be masked, only about 10 channels could be used for the analysis. This has a significant impact, since for many strips the charge collected in neighbouring channels cannot be taken into account when the impact particle creates signals in more than one strip. Also if the particle hits in a masked channel, the signal in its neighbouring channel that is considered for readout can be just above threshold, but a large part of the signal is not measured. Both effects shift the signal distribution and thus the MPV to a smaller value.  This is also having a minor impact on the collected charge of the \textit{low dose 30} design, as there also less than 20 channels were available for the analysis. 
Within the \textit{regular} and the \textit{low dose 30} designs there is no significant difference visible between the individual stitches. For the \textit{low dose 55} design there is a small difference visible, but with a higher charge for the stitch further away from readout, thus it is not considered to be an effect due to the stitching.

\begin{figure}[ht]
    \centering
    \includegraphics[width=0.75\linewidth]{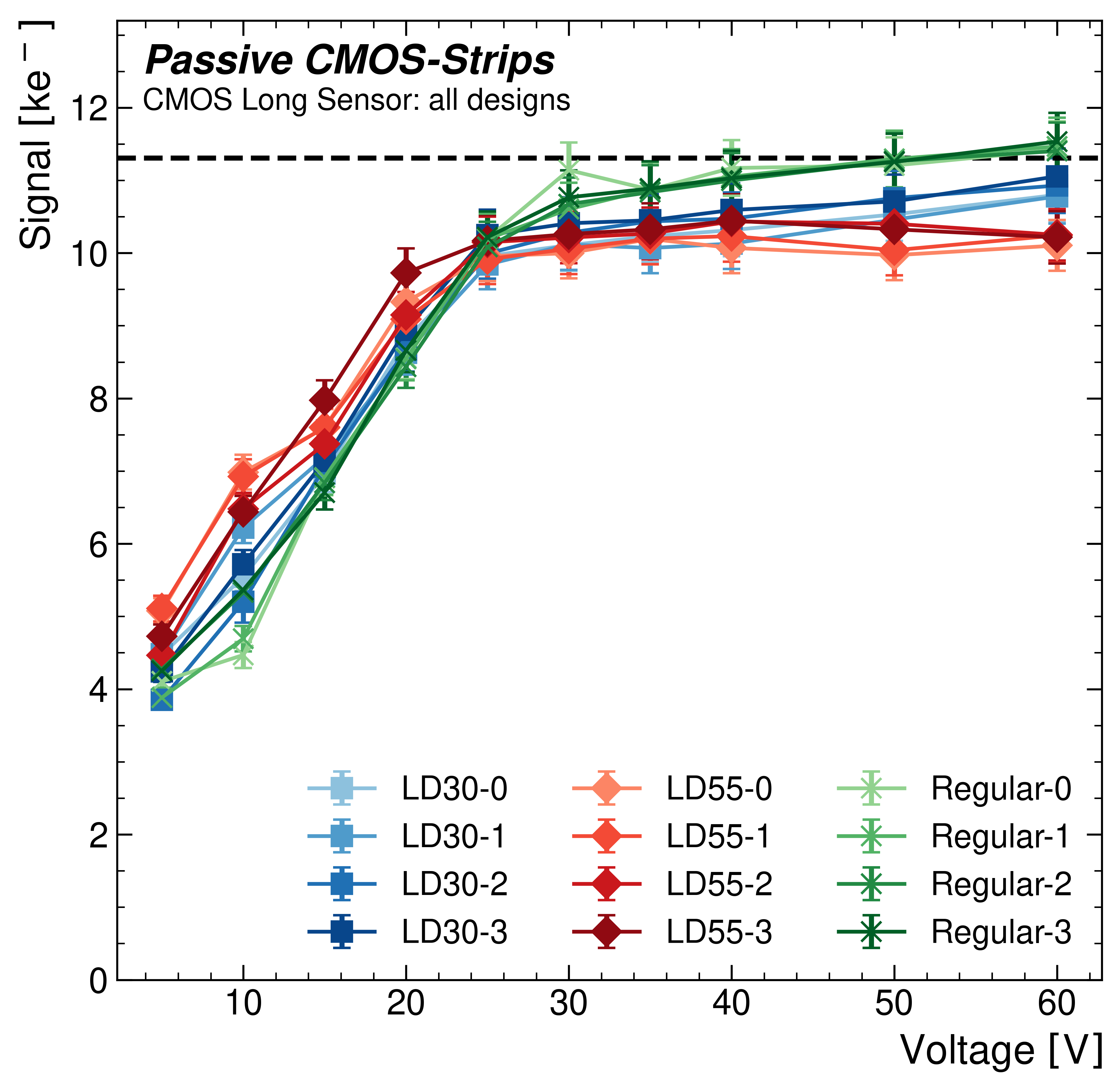}
    \caption{\it \small Signal as a function of the bias voltage in different stitched areas of a long p-CMOS-strip sensor. The stitching areas are numbered 0,1,2 and 3 as indicated in the label names (e.g. LD30-0 means \textit{low dose 30} stitch 0).
    \label{fig:long sensor signal}}
\end{figure}

This was confirmed also for the long sensors, where even four stitched areas, numbered 0-3, were measured. The collected charge of a long sensor as a function of the bias voltage is presented in Figure \ref{fig:long sensor signal} for the different designs and stitching areas. 
As for the short sensors, the depletion voltage of about 35~V is clearly visible. However, for this sensor only the \textit{regular} design measurements reach the expected amount of collected charge. Here the effect of the few channels usable for the analysis is visible for both \textit{low dose} designs. 

For the four stitches there is no difference in the course of the collected charge recognizable, thus also more stitch lines have no negative effect on the collected charge. 

\section{Laser Measurements}

Laser measurements enable a position dependent measurement of the collected charge, and thus are an important tool to investigate new sensor techniques. 
Two different types of laser measurements were performed. One is scanning across and along the edge of the sensor to prove that the full sensor thickness is depleted and no stitching effect along the edge are visible. The other one is scanning along the strips with a laser directed on the sensor top or the sensor bottom in order to evaluate the inter-strip region and if the stitch lines themselves show any effect on the collected charge.

\begin{figure}[ht]
    \centering
    \includegraphics[width=0.7\linewidth]{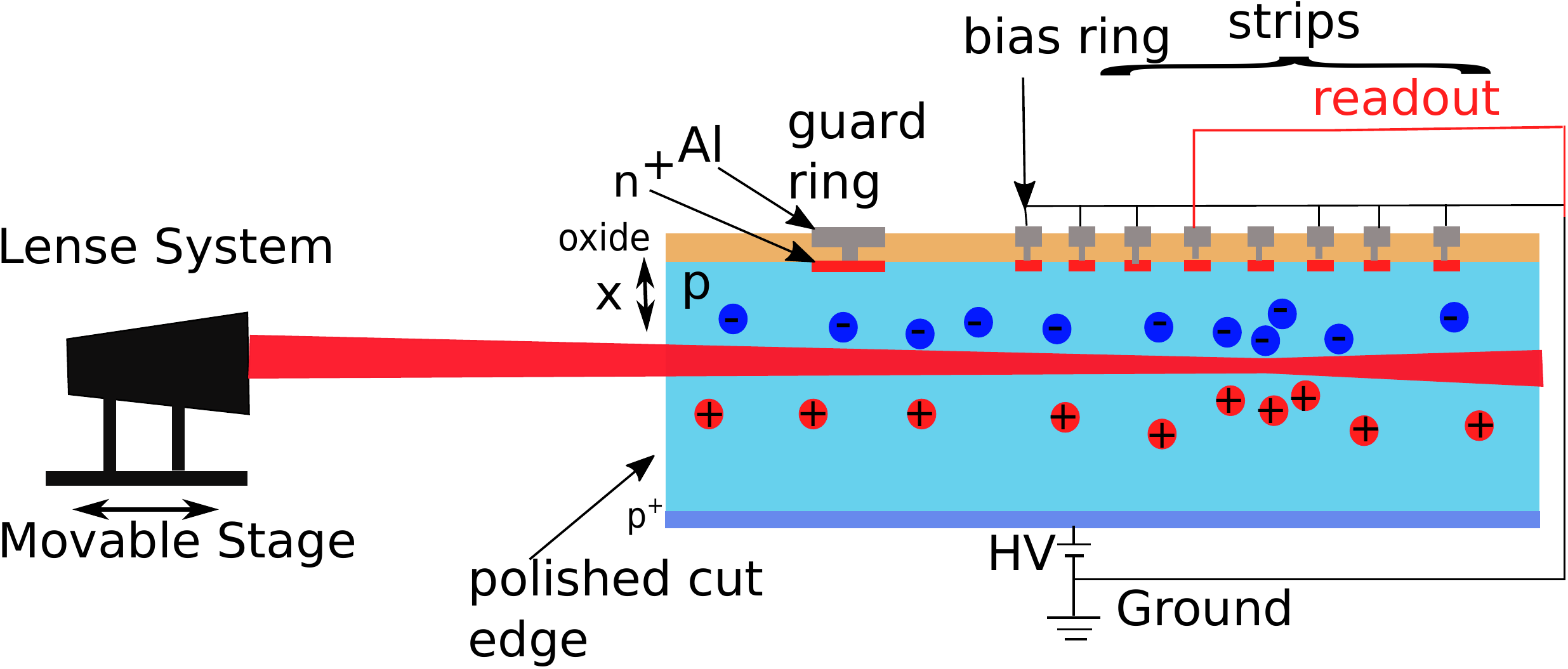}
    \caption{\it \small Measurement principle for edge-TCT measurements \cite{diehl2020effects}.
    \label{fig:Edge_Setup}}
\end{figure}

In order to evaluate the depletion depth in dependence of the voltage, edge-TCT (Transient Current Technique) measurements \cite{Kram1} were performed. 
The principle of the edge-TCT setup is sketched in Figure~\ref{fig:Edge_Setup}. A laser with a wavelength of 1064~nm creates electron-hole pairs at a specific depth, where the laser intensity is tunable between the equivalent of a few MIPs and about a few tens of MIPs. The laser beam has a Gaussian profile with a focus point width of below $10~\mu\mathrm{m}$, positioned beneath the strip connected to readout. The beam is directed perpendicular to the strips onto the polished cut-edge of the sensor. The readout strip is connected to an oscilloscope via a 53~dB amplifier while the neighbouring strips are connected to ground. The pulses are recorded and and the collected charge is calculated by integrating over the entire signal pulse, using usually a time window of 25~ns. 

\begin{figure}[ht]
    \centering
    \includegraphics[width=0.98\linewidth]{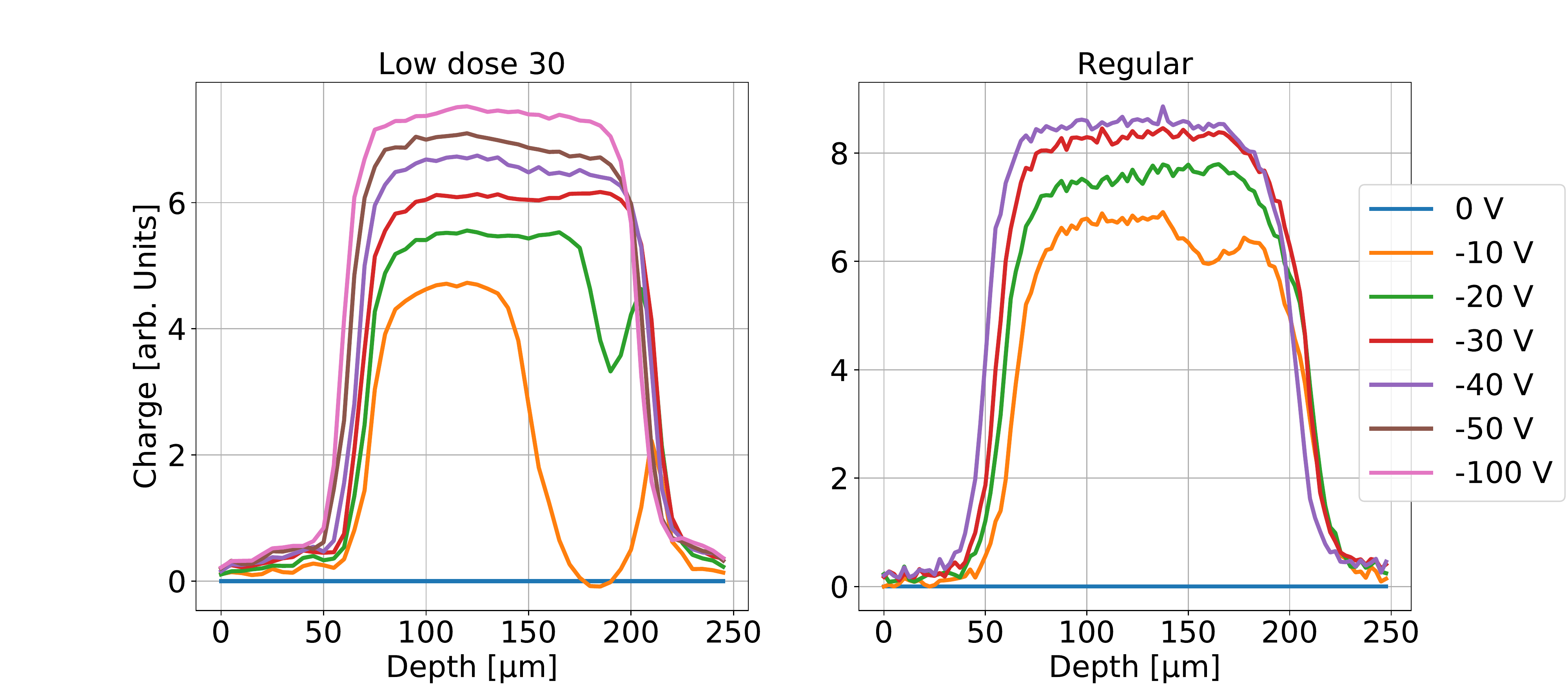}
    \caption{\it \small Collected charge as a function of depth for a short sensor for the low dose 30 and the regular design for several voltages.
    \label{fig:CC_vsDepth}}
\end{figure}

In Figure~\ref{fig:CC_vsDepth} the charge collection as a function of depth is shown for a short sensor for the \textit{low dose 30 design} (left) as well as the\textit{ regular} design (right) for several voltages. The \textit{low dose 55} design is not accessible via edge-TCT since it is too far away from the sensor edge. The voltage dependence is clearly visible by the increase of the depletion depth. 
The \textit{low dose 30} design reaches full depletion at around 30~V, however the charge collection keeps increasing slightly after the full sensor is depleted. The width of the depletion zone reaches $150~\mu\mathrm{m}$, confirming the nominal thickness of the sensors and proving that the entire sensor volume is sensitive to charge.
Also the \textit{regular} design shows a depletion voltage around 30~V, which agrees with the bulk capacitance measurements. Comparing the two designs, the \textit{regular }design seems to have a larger depletion depth at lower voltages, which might hint to slightly different electric field distributions while the sensor is not fully depleted. Both designs deplete from front to back as expected and the measurements are in agreement with the TCAD simulations of the electric field.

\begin{figure}[ht]
    \centering
    \includegraphics[width=0.85\linewidth]{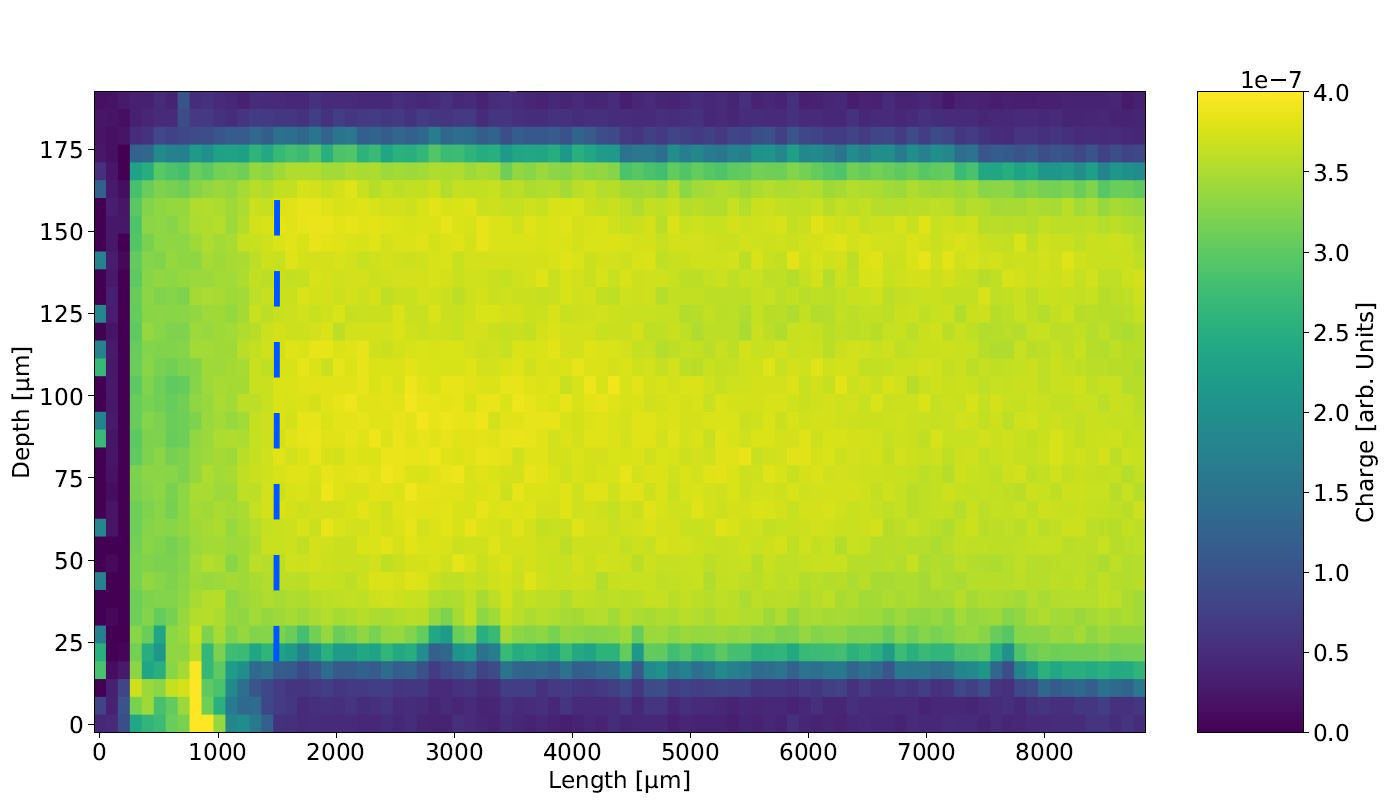}
    \caption{\it \small Collected charge as a function of depth and position along the strip for a short sensor for the \textit{low dose 30} design measured at 80~V.
    \label{fig:CC_vsDepth_vslength}}
\end{figure}

To investigate if the stitches have an impact on the charge collection, another measurement scanning across the sensor depth and along the edge was performed. 
In Figure~\ref{fig:CC_vsDepth_vslength} the collected charge as a function of sensor depth and position along the strip length is shown for the \textit{low dose 30} design of a short sensor at 80~V, including a sketch of the approximate stitch line position.  Along the sensor length a very uniform charge distribution is visible and no effect of the stitch line can be recognised. The few small areas with no charge collection at the bottom of the sensor, e.g. around $3000~\upmu\mathrm{m}$ are explained with glue remains blocking the laser light. For a length below $1000~\upmu\mathrm{m}$ there is some charge measured outside of the expected sensor area due to reflections on the metal box, metallisations of the PCB and the wire bonds on top of the sensor.

These laser measurements  illuminating the sensor from top or bottom are carried out using the ALIBAVA-readout system described in section \ref{sourcemeasuremnts}. Instead of a radioactive source, a laser beam with adjustable power, a wavelength of 980~nm and a focus width of at least $3~\upmu\mathrm{m}$ illuminates the sensor. The beam penetrates around $100~\upmu\mathrm{m}$ deep into the silicon.

\begin{figure}[ht]
    \centering
    \includegraphics[width=0.85\linewidth]{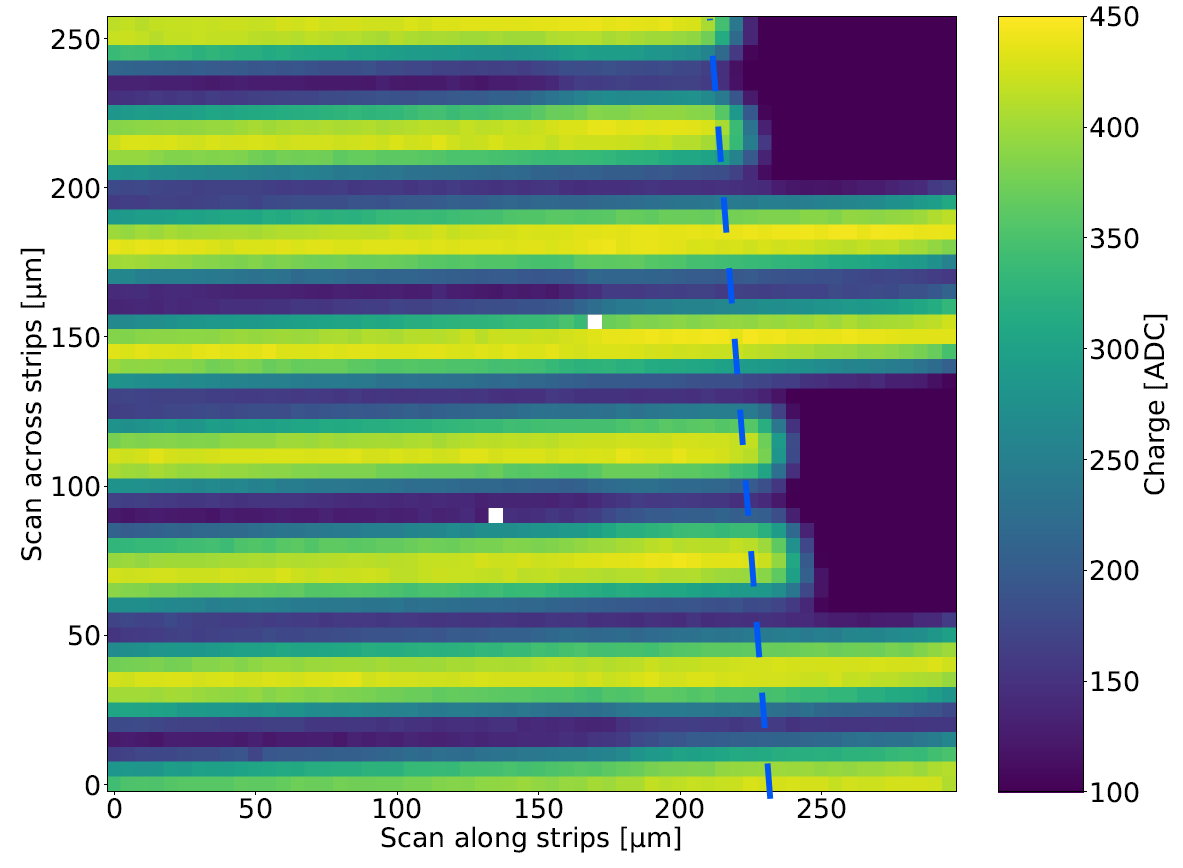}
    \caption{\it \small Collected charge of the\textit{ regular} design of a long sensor as a function of the laser position at 50~V, illuminating the sensor from the top.
    \label{fig:ALIBAVA_laser_top}}
\end{figure}

In Figure~\ref{fig:ALIBAVA_laser_top} the charge collection of the \textit{regular} design of a long sensor at $50~\mathrm{V}$ is presented as a function of the laser position. The sensor was illuminated from the top, thus there is no charge collected on the strips, bond pads and the interstrip metallisation, since the laser light is reflected.
The measurement extends across four strips in $5~\upmu\mathrm{m}$ steps. The four strips are clearly visible and as sketched, there is a stitch line directly beneath the bond pad area. Thus, the measurement clearly shows that stitching works and no effects are visible around the stitch line. 

\begin{figure}[ht]
    \centering
    \includegraphics[width=0.85\linewidth]{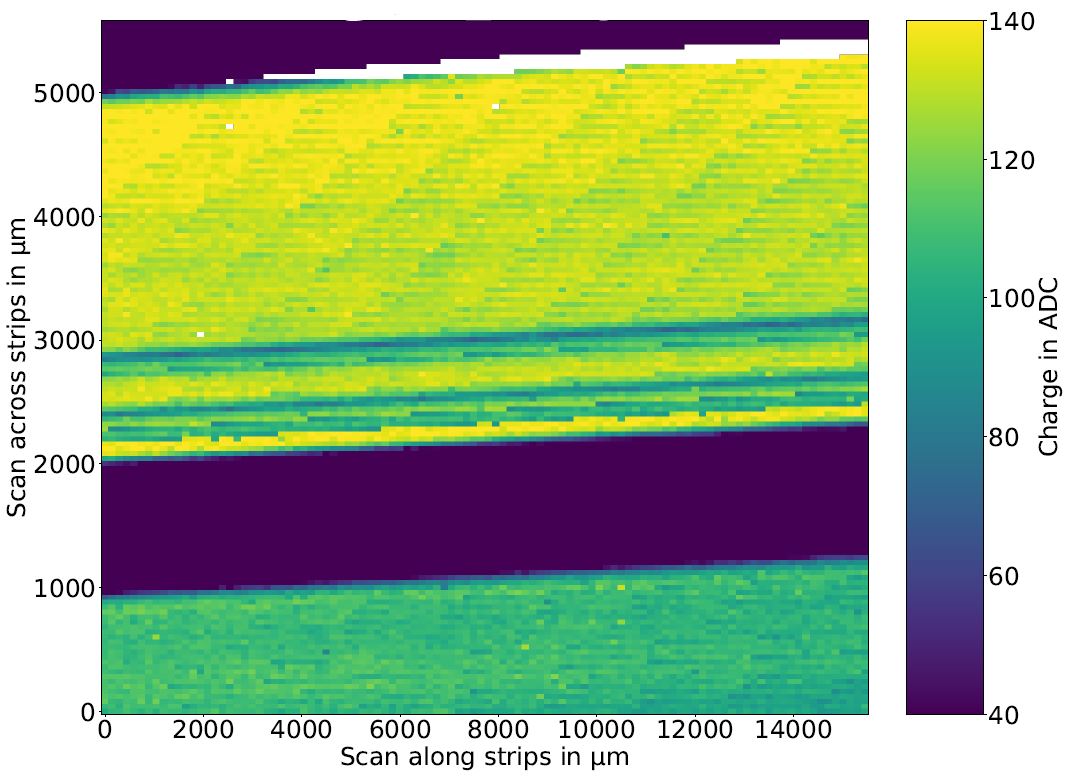}
    \caption{\it \small Collected charge  of a long sensor as a function of the laser position at 35~V, illuminating the sensor from the back.
    \label{fig:ALIBAVA_laser_back}}
\end{figure}

In the ALIBAVA laser measurements all strips are connected to readout, therefore larger areas of the sensor can be compared.
Since the sensors have no backplane metallisation, this was utilized to measure them without having reflections on e.g. the strip metallisation. 

In Figure \ref{fig:ALIBAVA_laser_back} the collected charge of a long sensor measured at 35~V illuminating the sensor from the back is presented for the \textit{regular} design (top) and the \textit{low dose 55} design (bottom). Since the sensor was not aligned perfectly parallel to the x-axis of the measurement set-up, the sensor is tilted a few degrees.
As can be seen, there is no effect from stitching visible as well, the collected charge in the two designs varies only 4\% each throughout the entire measured areas. In the \textit{regular} design it shows that four strips are not giving signal, which was caused by lost bonds. Between the two designs is an area without any collected charge, where no strips are connected to the ASIC readout channels. 
The two designs show a difference in charge of about 25 ADC (Analog to Digital Counts). Due to the different interstrip capacitances of the designs, the load on the Beetle chip is different and the calculation into k$e^-$ varies.  The measurements presented in section \ref{sourcemeasuremnts} prove that the collected charge in k$e^-$ of the different designs is actually approximately the same. 

\section{Summary and Conclusion}
Strip sensors produced in passive CMOS technology and employing stitching to connect several reticles, resulting in strips of up to 4 cm length, have been studied in detail. The results show that stitched passive CMOS sensors collected the expected amount of charge and the entire sensor volume is sensitive to charge. Especially, no negative effects arise from the stitching process. 
The electrical tests of the first batch produced showed a low breakdown voltage due to a non-optimal backside processing and missing backplane metallisation, which was corrected then for the second production batch. 
The charge collection of the three investigated designs is in agreement with the  $11.3~\mathrm{k}e^-$ in $150~\upmu\mathrm{m}$ silicon, with no measurable decrease of charge along the strips through the stitches. However, the low number of channels in the two \textit{low dose} designs affects the measurement of the collected charge negatively.  Possible differences in the performance of the designs are expected to arise after irradiation, thus conclusions about the best design choice cannot be drawn yet.
Laser measurements proved that the full active depth of $150~\upmu\mathrm{m}$ of the sensors depletes as desired, and there was no effect visible around the stitch lines. 
In conclusion, the CMOS process is a promising, cost-effective candidate for future experiments needing to cover large areas with silicon detectors for tracking. 
In further studies the radiation hardness, which is a key parameter especially  for sensors used in hadron collider experiments, will be under investigation as well.

\bibliography{mybibfile}

\end{document}